\documentclass[reprint, superscriptaddress, longbibliography, floatfix, twocolumn, nofootinbib, amsmath, amssymb,
 aps, showkeys]{revtex4-2}
\usepackage{mathtools}
\usepackage{nicefrac}
\usepackage{braket}
\usepackage{siunitx}
\usepackage{hyperref}
\usepackage{graphicx}
\usepackage{dcolumn}
\usepackage{bm}
\usepackage{dsfont}

\makeatletter
\newcommand*{\rom}[1]{\expandafter\@slowromancap\romannumeral #1@}
\makeatother

\begin{document}

\preprint{APS/123-QED}

\title{Demonstration of quantum network protocols over a 14-km urban fiber link}

\author{Stephan Kucera}
\thanks{}
\affiliation{Experimentalphysik, Universit\"at des Saarlandes, 66123 Saarbr\"ucken, Germany}
\affiliation{Luxembourg Institute of Science and Technology (LIST), 41 rue du Brill, L-4422 Belveaux, Luxemburg}

\author{Christian Haen}
\thanks{} 
\affiliation{Experimentalphysik, Universit\"at des Saarlandes, 66123 Saarbr\"ucken, Germany}

\author{Elena Arensk\"otter}
\thanks{} 
\affiliation{Experimentalphysik, Universit\"at des Saarlandes, 66123 Saarbr\"ucken, Germany}

\author{Tobias Bauer}
\thanks{}
\affiliation{Experimentalphysik, Universit\"at des Saarlandes, 66123 Saarbr\"ucken, Germany}

\author{Jonas Meiers}
\thanks{}
\affiliation{Experimentalphysik, Universit\"at des Saarlandes, 66123 Saarbr\"ucken, Germany}

\author{Marlon Sch\"afer}
\thanks{}
\affiliation{Experimentalphysik, Universit\"at des Saarlandes, 66123 Saarbr\"ucken, Germany}

\author{Ross Boland}
\thanks{}
\affiliation{Menlo Systems GmbH, Bunsenstraße 5, 82152 Planegg, Germany}

\author{Milad Yahyapour}
\thanks{}
\affiliation{Menlo Systems GmbH, Bunsenstraße 5, 82152 Planegg, Germany}

\author{Maurice Lessing}
\thanks{}
\affiliation{Menlo Systems GmbH, Bunsenstraße 5, 82152 Planegg, Germany}

\author{Ronald Holzwarth}
\thanks{}
\affiliation{Menlo Systems GmbH, Bunsenstraße 5, 82152 Planegg, Germany}

\author{Christoph Becher}
\thanks{}
\email{christoph.becher@physik.uni-saarland.de}
\affiliation{Experimentalphysik, Universit\"at des Saarlandes, 66123 Saarbr\"ucken, Germany}

\author{J\"urgen Eschner}
\thanks{}
\email{juergen.eschner@physik.uni-saarland.de}
\affiliation{Experimentalphysik, Universit\"at des Saarlandes, 66123 Saarbr\"ucken, Germany}

\date{\today}

\begin{abstract}	
We report on the implementation of quantum entanglement distribution and quantum state teleportation over a 14.4\,km urban dark-fiber link, which is partially underground, partially overhead, and patched in several stations. We characterize the link for its use as a quantum channel and realize its active polarization stabilization. Using a type-II cavity-enhanced SPDC photon pair source, a $^{40}\mathrm{Ca}^{+}$ single-ion quantum memory, and quantum frequency conversion to the telecom C-band, we demonstrate photon-photon entanglement, ion-photon entanglement, and teleportation of a qubit state from the ion onto a remote telecom photon, all realized over the urban fiber link.
\end{abstract}


\maketitle

Quantum networks require distributed entanglement as a resource \cite{Kimble2008, Wehner2018}, in order to enable elementary communication routines like the quantum repeater \cite{Briegel1998} or quantum state teleportation \cite{Bennett1993}, or more advanced protocols such as entanglement-assisted clock synchronization \cite{Shi, Jozsa}, phase-coherent frequency comparison of optical clocks \cite{Schnatz}, or distributed quantum sensing \cite{Zhang, Pelayo_2023}. This renders the specifications for quantum communication much stricter than for classical communication, especially when dealing with single photons and polarization-encoded qubits, that require low background and high polarization stability. It also defines the requirements for integrating quantum communication into existing communication networks using telecom fibers. 

Several proof-of-principle demonstrations have extended quantum network functionality from laboratory scale to real-world distances, prominently for quantum key distribution (QKD) over deployed fibers, e.g. \cite{Shi2020, Wengerowsky2018, Wengerowsky2020, Neumann2022, Chen2021, Ribezzo2022, zahidy2023, Pelet2023}. These demonstrations have addressed some of the challenges of integrating quantum communication into existing communication infrastructure, such as wavelength-division multiplexing (WDM) and co-propagation of classical data traffic and quantum signals on the same channel \cite{Clark_2023, Chung2022, Berrevoets2022, Avesani2022, lord2023, martin2023, wang2023}. More advanced protocols for quantum communication networks employ quantum memories, for example for quantum repeater operations \cite{Sangouard2009}. Demonstrations on laboratory scales have used memories based on single ions \cite{Krutyanskiy2023}, single atoms \cite{Weinfurter, Daiss2021}, solid-state qubits \cite{Pompili2021, Lago2023}, or atomic ensembles \cite{Rivera2021, Liu2021}. Some of them link different buildings on a campus, or include fiber-spools to simulate large distances. The implementations in \cite{Du2021, Luo2022, Liu2023, rakonjac2023, knaut2023} demonstrate basic functionality on a deployed fiber network using atomic ensembles, multimode solid-state quantum memories, or silicon-vacancy centers in nanophotonic diamond cavities.

The deployment of existing telecom fiber infrastructure for quantum communication protocols is a key step towards efficient development of quantum networks. It also entails multiple technical challenges, since existing fiber infrastructure in an urban region is often underground, or paired with the electrical overhead power line. In the latter situation, the fibers are installed in the earth wire, which is usually grounded. This wire serves to minimize the likelihood of direct lightning strikes to the other installed conductors \cite{Uman2008} but is not used for power transmission, therefore no Joule heating is expected. Nevertheless, environmental influences such as wind-induced mechanical stress \cite{Ding_17}, rain, ice, snow, and birds cause fluctuations in polarization mode dispersion, photon travel times \cite{Brodsky}, and polarization-dependent loss \cite{GISIN}. 

The figures of merit of optical fibers for classical communication are mainly their transmission and polarization mode dispersion (PMD). To characterize a channel for the transmission of polarization-encoded quantum bits, PMD must be investigated in more detail. It involves properties such as polarization, phase and travel duration in a time dependent or fluctuating manner. Such a characterization was carried out for the Boston-area quantum network testbed \cite{bersin2023}. Here, we present the exploration of a $14.4$\,km long deployed fiber link in the urban area of Saarbr\"ucken, Germany. The link connects the campus of Saarland University (UdS)\footnote{Campus E2 6, 66123 Saarbrücken, Germany} with the campus of the University of Applied Sciences (HTW)\footnote{Goebenstraße 40, 66117 Saarbrücken, Germany}. It comprises $1278$\,m of overhead cable and several patch stations. An efficient polarization stabilization scheme is implemented following the method of \cite{DiplomHocke}, to ensure high-fidelity transmission of photonic polarization quantum bits over the link. On the hardware side, we utilize a single trapped $^{40}$Ca$^+$ ion as quantum memory, an ion-resonant, high-brightness, high-fidelity spontaneous parametric down-conversion (SPDC) source of entangled photon pairs, and polarization-preserving quantum frequency conversion (QFC) from the ionic wavelength to the telecom C-band \cite{Arenskoetter2023, Arenskoetter2024}. As examples of quantum communication protocols, photon-photon and ion-photon entanglement distribution with photon pairs, as well as quantum-state teleportation from a single ion to a remote telecom photon are demonstrated over the fiber link.

\section{Fiber link as quantum channel}\label{fiberlink}
\begin{figure}[t]
\includegraphics[width=0.45\textwidth]{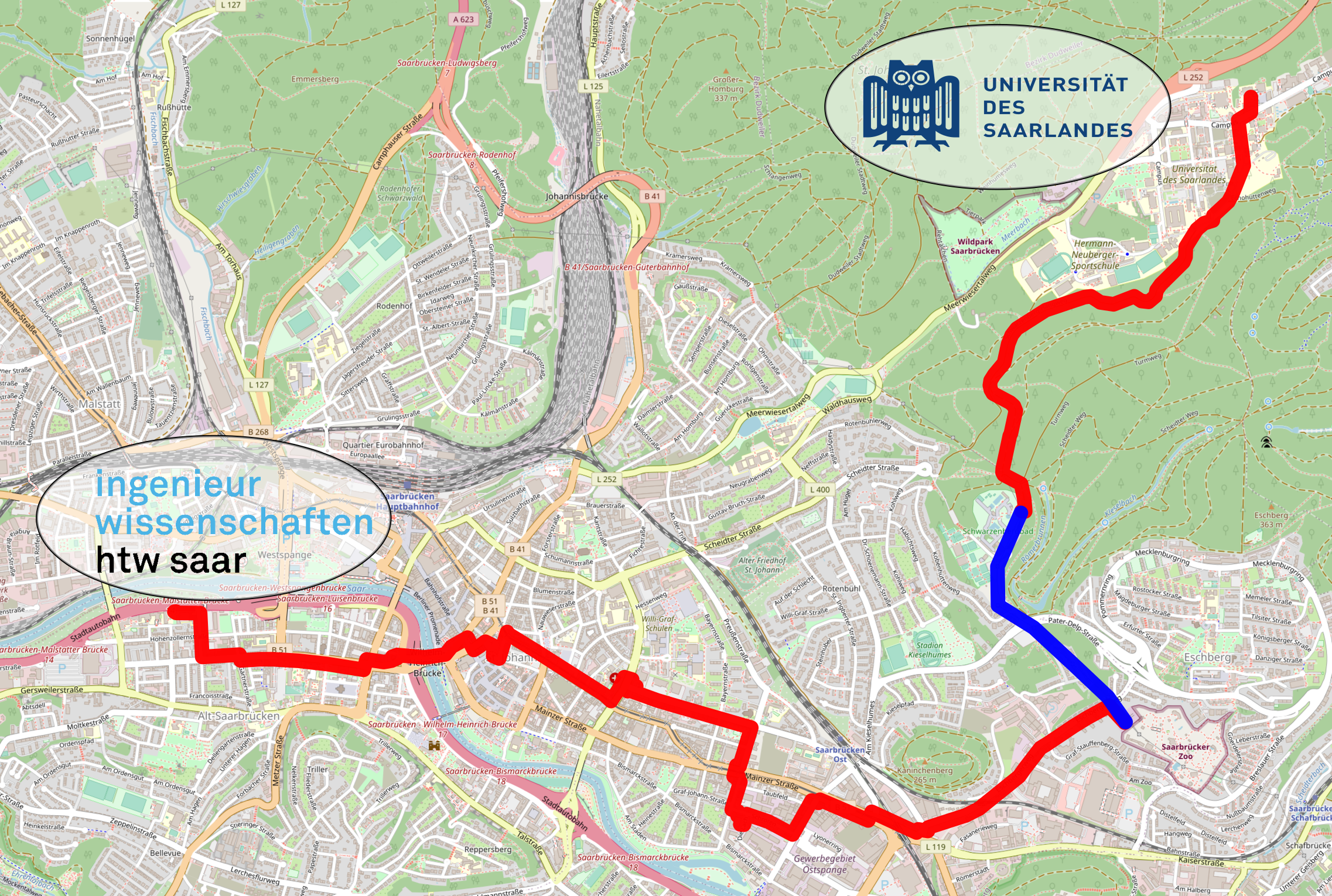}
\caption{Map of the $14350$\,m-long fiber link extending from the labs at the UdS to the HTW. Underground sections are in red, overhead sections in blue (map data from OSM \cite{OSM}).}
\label{fig:FiberLinkMap}
\end{figure}

The fiber is supplied by a local provider (VSE NET), its course across Saarbr\"ucken is shown in figure~\ref{fig:FiberLinkMap}. It runs from the HTW through the city center via underground cables (red lines) to a voltage transformer station in the outer Saarbr\"ucken region. From this station, it follows the overhead power line (blue line) for $1278$\,m and then continues underground through a forested area to a server room at the UdS. The university IT department (HIZ) provides the local fiber links at the HTW and the UdS between access points and laboratories. Two dark fibers from a deployed fiber bundle are available which will be denoted as Link-Q and Link-C, due to their use in the experiments as quantum and classical link, respectively. 

\subsection{Optical fiber loss}
\begin{figure}[t]
\includegraphics[width=0.4\textwidth]{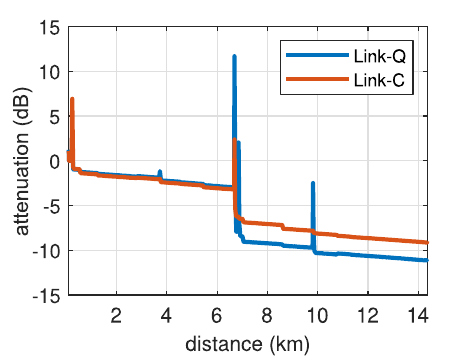}
\caption{OTDR measurement of the two fiber links from the UdS labs to the HTW.}
\label{fig:OTDR}
\end{figure}

The optical time-domain reflectometer (OTDR) measurement at 1550\,nm, shown in figure~\ref{fig:OTDR}, reveals a total loss of $10.4$\,dB ($8.9$\,dB) for the fiber Link-Q (Link-C). The fiber with higher loss is used as the quantum channel in the presented experiments. This has historical reasons and has meanwhile been changed.
The OTDR shows several patches as points of additional attenuation and a faulty splice at $6.6$\,km of $4.1$\,dB ($2.8$\,dB) loss in the fiber Link-Q (Link-C). The bare fiber attenuation is in the expected range between $0.19\,$dB/km and $0.23$\,dB/km.

\subsection{Light pollution}\label{sec:LightPollution}
\begin{figure}[t]
\includegraphics[width=0.4\textwidth]{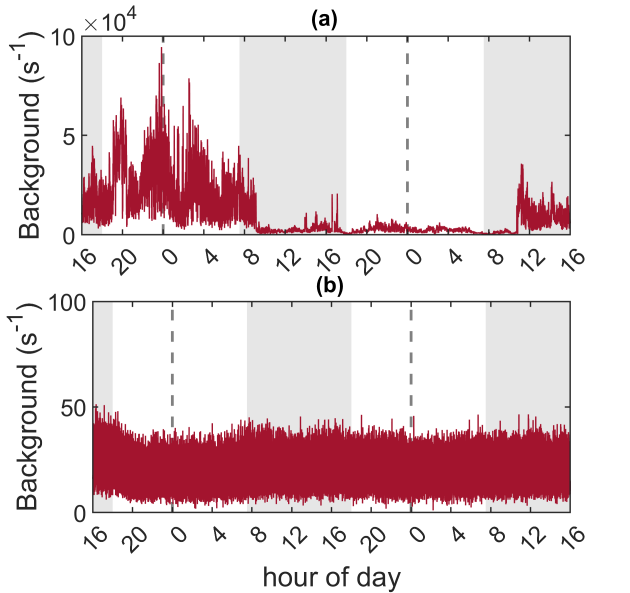}
\caption{Light pollution measurement: (a) fiber connected directly to the detector; (b) with $250$\,MHz bandpass filter installed. Note the change in the vertical scale between (a) and (b). }
\label{fig:background}
\end{figure}

Light pollution background, for example originating from sunlight or evanescently coupled light from neighboring optical fibers, is a major concern for experiments with single photons. Figure~\ref{fig:background}(a) shows a background measurement carried out with superconducting nanowire single-photon detectors (SNSPDs, detection efficiency $>80$\,\% at 1550\,nm). 
The large observed fluctuations are not correlated with day- or nighttime. We attribute these to background originating from activity at the patch stations or data traffic in the nearby located fibers. By design, the optimum detection efficiency for the detectors is at 1550\,nm, but still a broad spectrum of light is detected. To counteract the background, a broadband-suppression optical filter \cite{Arenskoetter2023} with a $250~$\,MHz transmission window is installed at the HTW in front of the detectors. The mean background level measured with this filter, shown in figure~\ref{fig:background}(b), is $19.7\,$s$^{-1}$ including detector dark counts. This filter is originally designed to suppress the optical noise induced by the quantum frequency conversion (see section \ref{sec:PDL}) but additionally also suppresses the environmental light.

\subsection{Polarization-dependent loss}\label{sec:PDL}
\begin{figure*}[t]
\includegraphics[width=\textwidth]{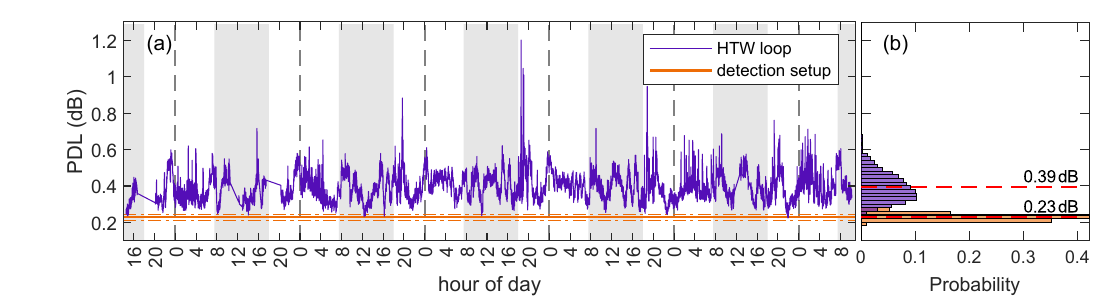}
\caption{Polarization-dependent loss. (a): time trace of total PDL of looped fiber link, $\mathcal{L}_\text{tot}$ (purple), including the detection device, recorded over around $6$ days, and PDL of the detection device alone, $\mathcal{L}_\text{det}$ (orange), with standard deviations (dashed lines), derived from a separate $22\,$h long measurement;
(b): histograms of PDL values $\mathcal{L}_\text{tot}$ and $\mathcal{L}_\text{det}$. We find expectation values for the PDL of $\mathcal{L}_\text{tot}=0.39(7)\,$dB for the full setup and $\mathcal{L}_\text{det}=0.23(2)\,$dB for the detection stage (red dashed lines). This results in $\mathcal{L}=0.08(9)\,$dB for the one-way fiber link.}
\label{fig:PDL}
\end{figure*}

Polarization-dependent loss (PDL) is of major importance for the transmission of polarization-encoded qubits. It is generally quantified by
\begin{equation}\label{eq:PDL}
	\mathcal{L}=10\log_{10} \left(\frac{T_\text{max}}{T_\text{min}}\right),
\end{equation}
where $T_\text{max}$ and $T_\text{min}$ are the measured maximum and minimum transmission values of a random set of polarizations \cite{Noe15}. We measured $\mathcal{L}_\text{tot}$, the PDL of the full fiber loop from the UdS to the HTW and back, by connecting the Link-Q and Link-C fibers at the HTW. We use two polarization scramblers, the first one consisting of a linear input polarizer and a set of motorized $\lambda/2$ and $\lambda/4$ waveplates to generate the initial polarizations, the second one using a piezo polarization controller (\textit{General Photonics PolaRITE\textsuperscript{TM} \rom{3}}) before the detection setup to decrease the influence of its own PDL on the measurement. In order to subtract the remaining PDL contribution of the detection setup alone, $\mathcal{L}_\text{det}$, we measured the transmission also without the fiber loop between the scramblers. The PDL of a single fiber is then inferred by $\mathcal{L} = (\mathcal{L}_\text{tot}-\mathcal{L}_\text{det})/2$, assuming uniform PDL for the two fibers.

A long-term measurement of $\mathcal{L}_\text{tot}$ is shown in figure~\ref{fig:PDL}(a). The orange baseline shows  $\mathcal{L}_\text{det}$. Polarization scrambling is performed by setting the first scrambler to 100 fixed polarizations across the Poincaré sphere and using 10 random polarizations at the second scrambler for each input. We calculate $\mathcal{L}$ according to eq. \ref{eq:PDL} by averaging over the 10 settings of the second scrambler and taking the maximum and minimum transmitted power of each input set. Each point of the time trace takes about $91\,$s to scramble the polarizations. In the time trace we see variation over the whole measurement time, but no correlation to weather or time of day. To quantify the PDL, we take the histograms of measured values shown in figure~\ref{fig:PDL}(b), and infer the single-fiber mean PDL of $\mathcal{L}=0.08(9)\,$dB. In appendix~\ref{appendix: PDL} a lower bound to the process fidelity of polarization qubit transmission, set by the PDL, is derived as $\mathcal{F}_{P} \geq \frac{1}{4}(1 + 10^{-\mathcal{L}/20)})^2$. For the measured mean PDL, $\mathcal{F}_{P} \geq 0.991$ is estimated. This fidelity is sufficiently high to neglect the influence of PDL on the polarization stabilization or on the quantum communication experiments.

\subsection{Polarization drift}\label{sec:poldrift}
To measure the capability of the fiber to transmit polarization qubits, we describe the quantum channel by a time-dependent real matrix $M(t)$ and an offset vector $\vec{c}(t)$, that act on the input Bloch vector $\vec{\lambda}_\text{in}$ to produce the output 
\begin{equation}
    \vec{\lambda}_\text{out} = M(t) \ \vec{\lambda}_\text{in} + \vec{c}(t) \ \ .
\end{equation}
This contains the full information of the quantum process on a single qubit \cite{Nielsen2000}. The amount of identity of the process, or process fidelity, is given by 
\begin{equation}
    \mathcal{F}_{\text{P}}(t) = \frac{1}{4}(1+\text{tr}(M(t)))~.
    \label{eq:ProcessFidelity}
\end{equation}
In our case, the low measured PDL value of $\mathcal{L}=0.08(9)$\,dB justifies that polarizing or depolarizing effects may be neglected, such that the offset vector $\vec{c}$ becomes zero and the matrix $M$ simplifies to a rotation matrix.

The description of the unitary polarization rotation $M$ for a single-photon polarization qubit, described by a Bloch vector $\vec{\lambda}$, is equivalent to that of classical polarization described by a Stokes vector $\Vec{S}$. Experimentally, the rotation matrix is determined by injecting two alternating reference lasers with fixed polarizations at the input and measuring the corresponding polarizations at the output. By using horizontal ($\Vec{S}_{H} = (1,0,0)$) and diagonal input polarizations ($\Vec{S}_{D} = (0,1,0)$), and measuring their output polarizations $\Vec{S}_1$ and $\Vec{S}_2$, the trace of the rotation matrix simplifies to
\begin{align}
    \text{tr}(M) &= \text{tr}([ \Vec{S}_1,\Vec{S}_2,\Vec{S}_1 \times \Vec{S}_2 ])\nonumber\\ 
    &= S_{1,1}+S_{2,2}+S_{1,1}S_{2,2}-S_{1,2}S_{2,1}\ .
    \label{eq:traceM}
\end{align}
 
\begin{figure*}[t]
\includegraphics[width=\textwidth]{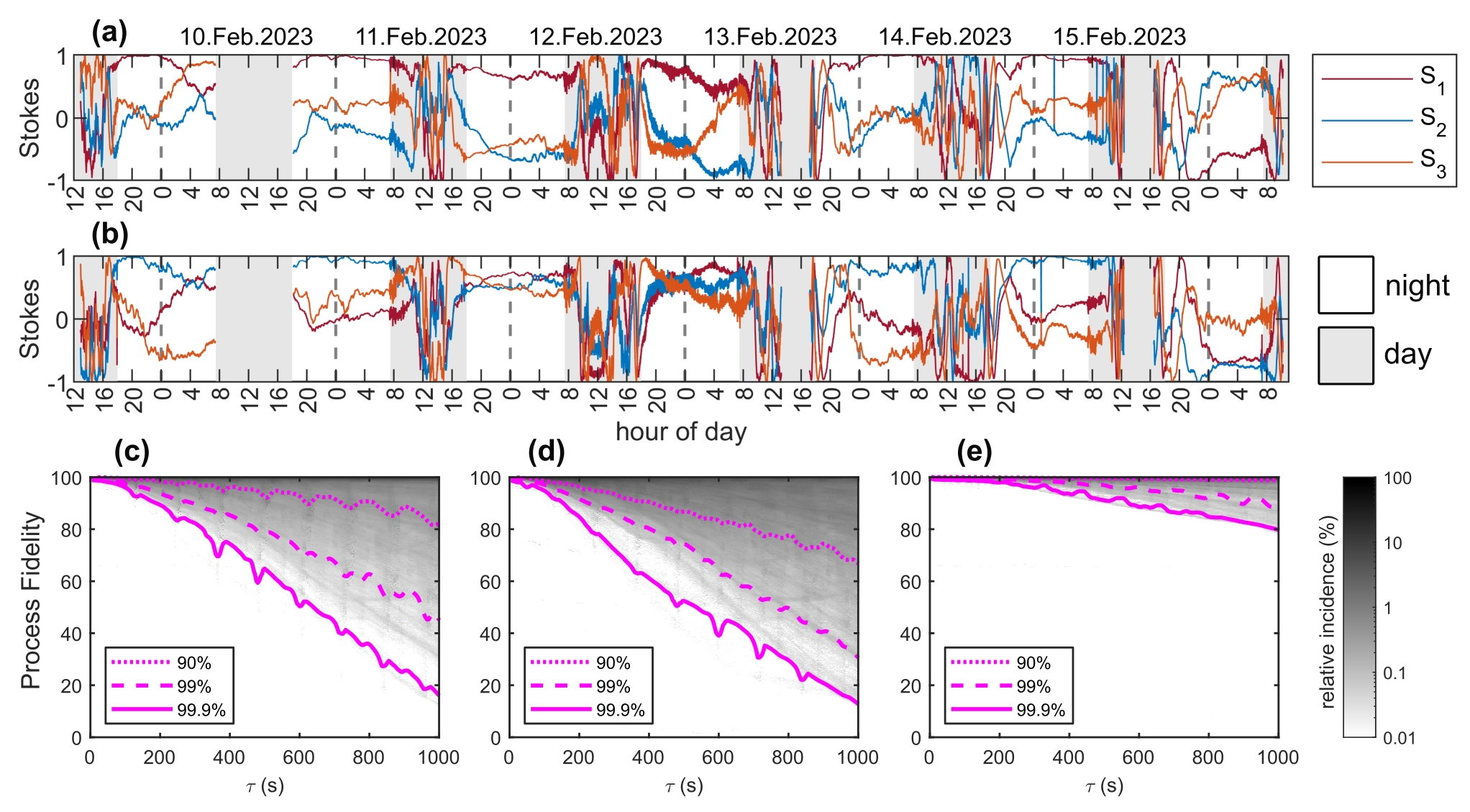}
\caption{Polarization drifts. (a) and (b): output Stokes parameters for input polarizations $\ket{H}$ and $\ket{D}$, respectively. (c), (d), and (e): measured probability (gray scale and violet lines) to maintain a certain process fidelity after a given free drift time $\tau$, evaluated for (c) the full time span, (d) for daytime only, (e) for nighttime only.}
\label{fig:StokesCollection}
\end{figure*}

A seven day-long measurement of the polarization drift is shown in figure \ref{fig:StokesCollection}(a-b). 
The gray-shaded areas highlight the daytime between 7:30 - 18:00, during which the fluctuations are visibly stronger than during nighttime. To quantify the time dependence, we determine the rotation matrix $M(t,\tau)$ that connects the output vectors with time separation $\tau$ according to $\Vec{S}_\text{i}(t+\tau) = M(t,\tau)\ \Vec{S}_\text{i}(t)$. From $M(t,\tau)$ we calculate the process fidelity using eqs.~(\ref{eq:ProcessFidelity}-\ref{eq:traceM}). The plots in figure \ref{fig:StokesCollection}(c-e) display the dependence of the process fidelity on the time difference $\tau$ for day and night (c), daytime only (d), and nighttime only (e). The gray-scale marks the relative incidence of measured process fidelities. The lines are the 99.9\% (solid), 99\% (dashed), and 90\% (dotted) quantiles. One sees that, depending on the requirements on the channel process fidelity, the channel can be used free-drifting for a certain period $\tau$ until a required threshold is reached and re-calibration or re-stabilization becomes necessary.

\subsection{Time spread}\label{sec:Timespread}
\begin{figure}[t]
\includegraphics[width=0.4\textwidth]{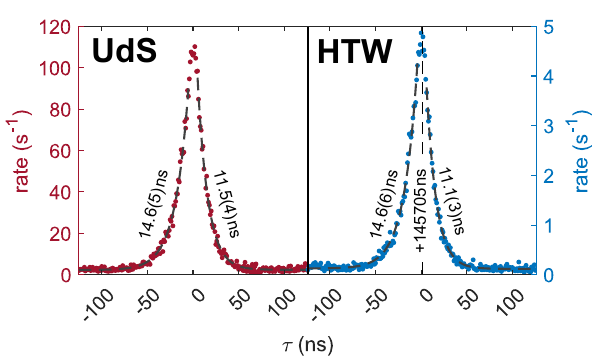}
\caption{Transmission time jitter of fiber link. Arrival time correlation of photon pairs generated at 854\,nm at the UdS and measured in the same laboratory (red, left side), and measured after one photon is frequency-converted and transmitted to the HTW via Link-Q (blue, right axis). The time axis at the HTW is centered at $+145,705$\,ns. The rates in $1$~ns-size bins are determined with $5\,$s ($80\,$s) integration time at the UdS (HTW). The dashed exponential fits reveal the time constants of the wave packets, indidated in the figure. }
\label{fig:wavepacket}
\end{figure}

The implemented quantum network experiments, described in section \ref{sec:QcomDemo}, require the detection times of the photons with a precision of $\approx1$\,ns, corresponding to a fraction of the atomic Larmor period (62.5\,ns). Fiber-induced time spread or jitter would diminish the fidelity of networking protocols. Therefore, two measurements were carried out to measure such effects. The first measurement uses the photon pair source of \cite{Arenskoetter2023} and evaluates the arrival time correlation ("wave packet") between photons of the same pair, with and without transmission through the fiber link. Figure~\ref{fig:wavepacket} displays the recorded correlation, when both photons are detected at the UdS (left), and when one photon is converted to 1550\,nm and transmitted to the HTW via the Link-Q fiber (right). (The digital signal of the SNSPD at the HTW is time-stamped at the UdS after is is sent back via the Link-C fiber; see section \ref{sec:PolDetHTW} for details.) By comparing the two time constants of the wave packets shown in \ref{fig:wavepacket}, we can bound the fiber-induced jitter to less than $600$\,ps over the 80\,s integration time. 

\begin{figure*}[t]
\includegraphics[width=\textwidth]{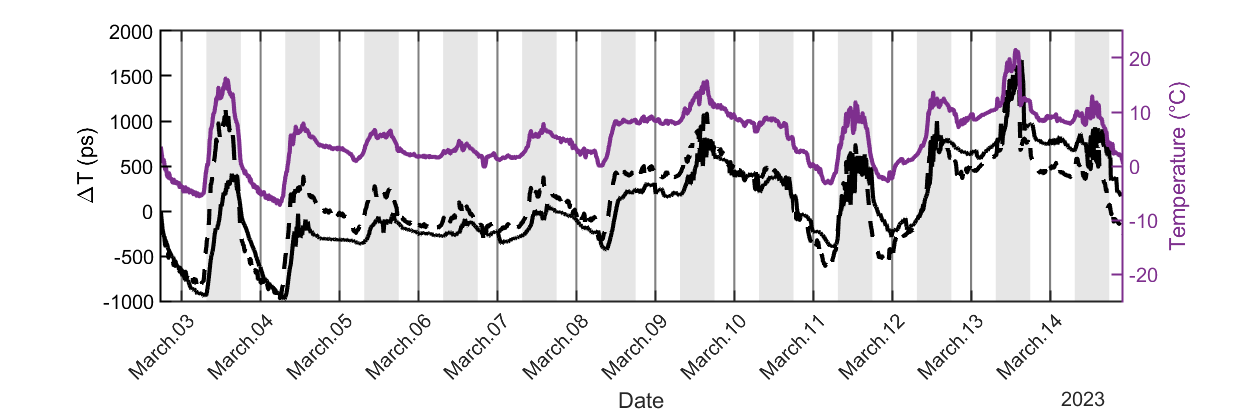}
\caption{Transmission time drift. The measured drift in the fiber link (solid black line) over 12 days closely matches the expected value (dashed black line) derived from the known temperature sensitivity of the SMF-28 fiber and the temperature data (purple line).}
\label{fig:TimespreadHIFI}
\end{figure*}

The second approach, implemented by Menlo Systems, uses light from a 1500\,nm auxiliary laser (\textit{NKT Koheras BASIK X15}), situated at the UdS, which is sent through the Link-C fiber. At the HTW, an acousto-optical modulator (AOM) shifts the laser frequency by a predetermined value, and a retroreflector sends the light back to the UdS. There, the beat signal, resulting from the interference between the local reference laser light and the returned light, is detected and analyzed using a frequency counter (\textit{FXE, K+K Messtechnik}). A change in the optical path length within the fiber induces a phase shift and, if time-dependent, a Doppler shift in the frequency of the light as it travels through the fiber \cite{MauriceLessingThesis}. The offset frequency due to the AOM is necessary to resolve the Doppler shift from the beat signal and also to distinguish the light reflected at the HTW from reflected light at splices and fiber connections. The Doppler shift is given by $\Delta \nu_\mathrm{D} = \frac{2}{c}\frac{\text{d}{nL}}{\text{d}{t}}\nu_0$, where $c$ is the speed of light in vacuum, $nL$ the optical path length of the fiber, $\nu_0$ the unshifted frequency of the light and the factor of 2 is due to the light passing the fiber twice. For each measured Doppler shift we derive the time delay $\Delta \text{T} = \frac{\Delta \nu_\mathrm{D}}{2\nu_0}t_\text{gate}$, where $t_\text{gate}=10\,\text{ms}$ denotes the sampling rate of the frequency counter. The fluctuating time delay measured over a period of 12 days is shown in figure \ref{fig:TimespreadHIFI} ($\Delta$T represented by the solid black line and averaged over 100 data points). The primary cause of this drift are the temperature fluctuations in the 2 × 1278\,m of overhead fiber; this is corroborated by the close correlation between the observed $\Delta$T values and the anticipated values (dashed black line) derived from the known temperature sensitivity of SMF-28 fibers (37.4\,$\frac{\text{p}s}{\text{km\,K}}$, as referenced in \cite{Slavik2015}) and the temperature data for the measurement period, taken from \cite{DWD} (purple line).

The implemented methodology allows us to resolve and also stabilize the optical path length for future investigations of quantum network protocols using optical interference of single photons, especially when deployed fiber infrastructure in overhead cables is used. 

\section{Experimental setup and polarization stabilization}

\begin{figure*}[t]
\includegraphics[width=\textwidth]{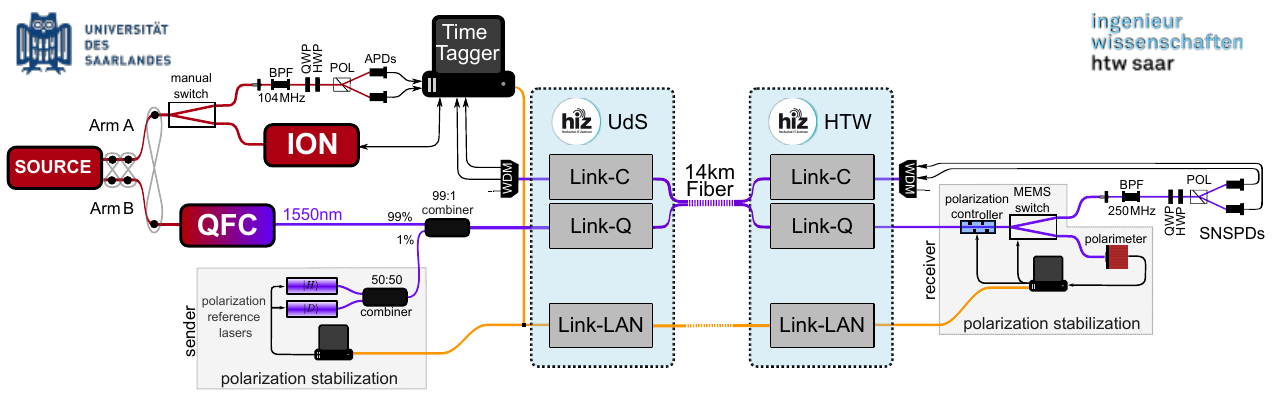}
\caption{Overall setup for the quantum communication protocols. Details are provided in the text. }
\label{fig:setup}
\end{figure*}

The overall experimental setup for realizing quantum communication protocols is shown in figure~\ref{fig:setup}. The main components are the photon pair source (SOURCE) with quantum frequency conversion (QFC), the ion-trap setup (ION), the fiber link itself, the polarization stabilization setup consisting of sender and receiver units, and polarization detection setups at the UdS and the HTW. The Link-LAN Ethernet connection serves for communication between the laboratories and for remote control of the HTW setup. In the following the components are described in more detail.

\subsection{Polarization detection at the HTW and Link-C connection}\label{sec:PolDetHTW}
The polarization detection setup at the HTW consists of a bandpass filter, a set of motorized quarter- and half-waveplate, and a Wollaston prism with one SNSPD at each output (\textit{Quantum Opus one}). The filter (BPF), centered around the frequency of the converted photons, serves to suppress the conversion-induced noise of the QFC device and the light pollution background (section \ref{sec:LightPollution}) outside its 250\,MHz transmission bandwidth \cite{Arenskoetter2023}. The projection basis is set by the waveplates, remotely controlled via the Link-LAN connection. The output pulses of the detectors are re-converted to optical pulses and transmitted via the Link-C fiber to the UdS, where they are again converted to electrical pulses and recorded by time-tagging electronics. The conversion processes are effected by RF-transmitters and receivers (\textit{ViaLite HRT-U1-6R-05-C/HRR-U1-6r-05}) and coarse wavelength division multiplexing (CWDM), using separate CWDM channels for each detector.

\subsection{Polarization stabilization}\label{sec:PolStab}
The polarization stabilization setup consists of a sender and a receiver unit. The sender unit features two switchable reference lasers, one with horizontal, the other with diagonal linear polarization. The two light sources are injected via a combination of a 50:50 and a 99:1 fiber beam splitter into the Link-Q fiber at the UdS and are alternately switched via ethernet. The receiver unit consists of a computer-operated piezo polarization controller (same as used in the PDL measurement) and a computer-controlled micro-electro-mechanical systems (MEMS) switch that directs the light either to a polarimeter (\textit{Thorlabs PAX1000IR2}) for stabilization, or to the detection setup (sec.~\ref{sec:PolDetHTW}), for communication. With the sender set to a fixed polarization, the time to measure the received polarization with the polarimeter and modify it via feedback to the polarization controller is about $90$\,ms. A total run of the polarization stabilization algorithm, explained in detail in appendix \ref{appendix:PolStabAlgorithm}, which includes $8$ feedback iteration, laser switching and calculation, takes $1.1(5)\,$s. Employing the MEMS switch, stabilization runs and quantum communication experiments are then performed in alternation.

Setting a targeted threshold fidelity $\mathcal{F}_\text{P,th}=99$\,\%, and starting from an arbitrary polarization rotation, the mean stabilization duration is measured as $t_\text{s}=6.4(17)$\,s. The time the algorithm needs starting from an initial process fidelity of $\mathcal{F}_\text{P}=96$\,\% is measured as $t_\text{r}=3.3(0.6)$\,s. This corresponds to the situation that after a preceding stabilization run and some free drift, a small detected rotation needs to be compensated. In the measurements described in section \ref{sec:QcomDemo}, we stabilize as soon as we detect a drift below $\mathcal{F}_\text{P}=99$\,\%, which on average takes only one iteration.
\begin{figure}[t]
\includegraphics[width=0.4\textwidth]{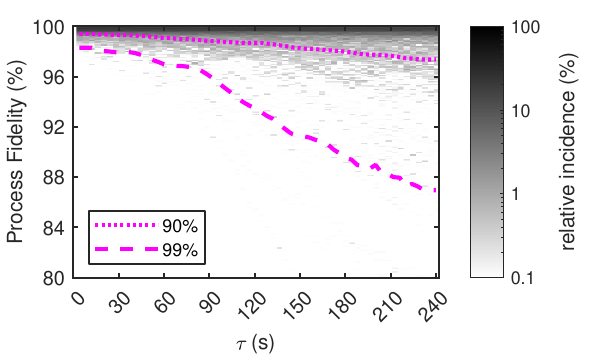}
\caption{Polarization drift after initial stabilization to a threshold process fidelity of $\mathcal{F}_\text{P,th}=99$\,\%.}
\label{fig:stabilisedPFsurfplot}
\end{figure}
Similar to section \ref{sec:poldrift}, we characterized the free polarization drift after a stabilization run. For that, the process fidelity after stabilization to $\mathcal{F}_\text{P,th}=99\%$ is recorded over a time $\tau$ of up to 240\,s. Figure \ref{fig:stabilisedPFsurfplot} shows the relative incidence (gray-scale) of measured process fidelities in dependence of $\tau$, evaluated from $450$ drifts. The lines are the 99\% (dashed), and 90\% (dotted) quantiles. The dependence is similar to the free polarization drift of figure \ref{fig:StokesCollection}. As an example, one reads from figure \ref{fig:stabilisedPFsurfplot} that with 90\,\% certainty the process fidelity is above $\mathcal{F}_\text{P}>98$\,\% for up to 176\,s, now including all the components of the fiber link and the stabilization.

\subsection{Photon pair source with quantum frequency conversion}\label{sec:photonpairsource}
The photon pair source, described in detail in \cite{Arenskoetter2023}, is based on cavity-enhanced spontaneous parametric down-conversion (SPDC) in an interferometric configuration. It produces the polarization-entangled two-photon state 
\begin{equation}
    \ket{\Psi}=\frac{1}{\sqrt{2}}\left(\ket{H}_\text{A}\ket{V}_\text{B}-e^{-i\phi}\ket{V}_\text{A}\ket{H}_\text{B}\right),
\end{equation}
with adjustable phase $\phi$. The photon in output arm\,A is resonant with the D$_{5/2}$-P$_{3/2}$ transition of $^{40}$Ca$^+$ at 854\,nm and and has 12.29\,MHz bandwidth, matching the 22\,MHz atomic transition linewidth. The photon in arm\,B is detuned to its partner by 480\,MHz. The latter is converted to 1550\,nm by a quantum frequency converter (QFC) based on difference frequency generation in periodically poled lithium niobate (PPLN) waveguides. Employing a Sagnac scheme, polarization independence of the conversion process is ensured. With a narrowband filter stage (250\,MHz FWHM) the conversion-induced noise count rate is reduced to $24\,\text{s}^{-1}$. Including the filters, the external device efficiency is 57.2\%.  Details on the converter are also presented in \cite{Arenskoetter2023}.

For the quantum communication experiments that are described in section~\ref{sec:QcomDemo}, the phase $\phi$ was set to $0^\circ$ to produce the maximally entangled $\ket{\Psi^+}$ Bell state. The pump power was set to $15\,$mW, generating fiber-coupled entangled photon pairs at $854$\,nm with a detected coincidence rate of $3352.0\,$s$^{-1}$ ($144.4\,$s$^{-1}$) when the partner photon is detected at the UdS (HTW), measured by integrating over the respective wave packet of figure \ref{fig:wavepacket}. The corresponding background rates are $2.32\,$s$^{-1}$ ($0.12\,$s$^{-1}$). 
The effective reduction of the coincidence rate between UdS and HTW amounts to $13.66\,$dB. This value results from the difference between $9.07\,$dB attenuation when a local projection and detection setup is inserted into arm B \cite{Arenskoetter2024}, and $22.73\,$dB total attenuation when the arm B photons are directed to HTW. The independently measured loss contributions to the latter are: QFC and transmission between the UdS labs ($6.78$\,dB), Link-Q fiber ($10.4$\,dB), transmission through the polarization stabilization sender ($0.46\,$dB) and receiver ($1.3\,$dB), filtering and projection setup ($0.65\,$dB) and detector efficiency ($0.97\,$dB) at the HTW. The remaining loss of $2.17\,$dB is attributed to fiber connections and optical misalignment at the remote HTW site during the time between setup and measurement.

\subsection{Ion trap setup and polarization detection at the UdS}
The photons in arm\,A  of the pair source are sent either to a polarization projection and detection setup, or to the ion trap setup. The polarization projection consists of a 134\,MHz bandpass filter, a set of motorized quarter- and half-wave plate, and a Wollaston prism with one APD at each output. The ion trap setup features a linear Paul trap with a single trapped $^{40}$Ca$^+$ ion and high-numerical-aperture objectives (NA$=0.4$) to address the ion with the 854\,nm photons and to collect emitted 393~\,nm photons. Laser and RF pulses enable coherent control of the initial state of the ion, as well as state readout after applying a quantum communication protocol, by projective measurement in a selected spin basis. The detectors of both setups are connected to time-tagging electronics. Further details on the ion trap setup and control are presented in \cite{Arenskoetter2024}.

\section{Quantum communication protocols}\label{sec:QcomDemo}

The demonstrated quantum communication protocols comprise (i) the distribution of entanglement provided by the photon pair source via the fiber link, (ii) the generation of distant ion-photon entanglement via quantum state-preserving heralded absorption of one photon of an entangled pair and transmission of the partner photon to the HTW, and (iii) quantum state teleportation from a memory qubit, encoded in the ion's spin state, onto a 1550-nm photonic qubit at the HTW. The teleportation protocol is described in detail in \cite{Arenskoetter2024}; here we focus on the results and the verification of the protocols as an application using the urban fiber link.

\subsection{Photon-photon entanglement}
To demonstrate the generation of distant photon-photon entanglement between the UdS and the HTW, the photons in arm\,A are sent to the projection setup while the photons in arm\,B are transmitted to the HTW via the Link-Q fiber after quantum frequency conversion to 1550\,nm. Quantum state tomography \cite{James2001} is then used to reconstruct the photon-photon density matrix. For the tomography, 16 basis combinations are measured. The process fidelity threshold of the polarization stabilization is set to $\mathcal{F}_\text{P,th}=99\,\%$. The polarization basis at the HTW is additionally calibrated every half hour, in order to correct for slow drifts in the unstabilized fibers from the SPDC source to the QFC device (90\,m), and from the QFC to the injection into the Link-Q fiber (100\,m). This calibration is applied in an automated way with single photons from the pair source: By changing its internal settings, it produces only horizontally polarized photons in arm B, which are alternately rotated to right-hand circular polarization with a motorized waveplate. These two input polarizations and the polarization projection setup at the HTW are used to measure the rotation matrix of the whole path, and the detection basis is adjusted accordingly. The same method is used every two hours to correct for drifts in the fibers that connect the unconverted photons of path A to the local $854\,$nm projection setup. The correction schemes are applied in all three protocols described here.
\begin{figure}[t]
\includegraphics[width=0.4\textwidth]{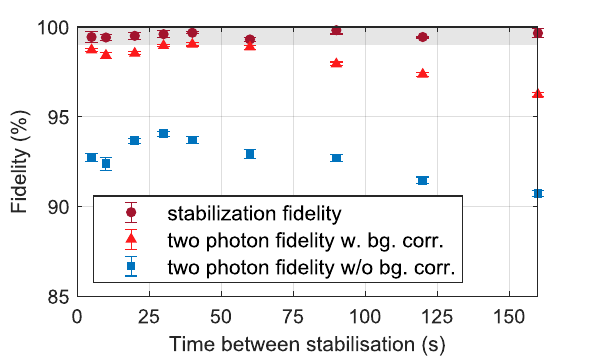}
\caption{Entanglement distribution. Photon-photon entanglement fidelity in dependence of time between two polarization stabilization runs with background correction (orange triangles) and without (blue squares). The error bars are calculated with Monte Carlo simulations \cite{ALTEPETER2005105}. The mean process fidelity at the end of each stabilization run (red dots) is always above the chosen threshold of $99\,\%$ (gray shaded area). The error bars are calculated by evaluating the standard deviation of all cycles at each point.}
\label{fig:PPEntWStabilization}
\end{figure}

The result of the photon-photon entanglement experiment is summarized in figure \ref{fig:PPEntWStabilization}, showing the relevant process fidelities when the time between polarization stabilization runs is varied from 5\,s to 160\,s. The red dots are the mean process fidelity of the polarization correction, $\mathcal{F}_\text{P}$, at the end of each stabilization run. The blue squares are the corresponding fidelities of the two-photon state to the maximally entangled Bell state $\ket{\Psi^+}$, computed from the reconstructed two-photon density matrix. The fidelities depicted by orange triangles include correction for background that is intrinsic to the SPDC process; details are explained in \cite{Arenskoetter2023}. With this correction, one can infer the quality of the fiber link alone, or the maximum pair fidelity one could reach with a photon source of larger purity. To acquire comparable statistics, the integration time for each basis is chosen to $\approx100\,$s and coincides with the time between the stabilization periods for the last three data points. We find qualitative agreement of the observed background-corrected fidelity ($\mathcal{F}>98\,\%$) with the result of section \ref{sec:PolStab}, where a slow decrease in fidelity is visible for larger waiting times between two stabilization runs. At times $>60$\,s, the influence of the fiber on the generated two-photon state becomes visible. For waiting times below 60\,s the implemented scheme is well suited to transmit the produced two-photon state without significant fidelity loss. Compared to the $>99\,\%$ background-corrected fidelity reported in \cite{Arenskoetter2023}, only a minor reduction due to the use of the deployed fiber was measured, which can be mostly attributed to polarization-dependent loss, discussed in section \ref{sec:PDL}.

\subsection{Ion-photon entanglement and ion-to-photon quantum state teleportation\label{sec:QcomDemoEnt}}
In these experiments, we distribute the entanglement of the photon pair source by sending one frequency-converted photon in the telecom C-band to the HTW and its partner photon to the trapped $^{40}$Ca$^+$ ion, which serves as a quantum memory. Two different protocols are applied, both based on state-preserving heralded absorption of a photon by the ion, presented in \cite{Kurz2014, Kurz2016, Arenskoetter2024}. In the first protocol, the photonic qubit from arm A is mapped onto the internal memory qubit of the ion, such that distant entanglement between the ion and the partner photon at the HTW is created. The second protocol starts with the preparation of a qubit state in the ion and uses the subsequent heralded absorption as a Bell-state measurement (BSM) on the ionic qubit and the polarization qubit of the absorbed photon \cite{Arenskoetter2024}. The result of the BSM facilitates teleportation of the prepared memory qubit onto the partner photon at the HTW. 

\begin{figure}[htb]
	\includegraphics[width=8.5cm]{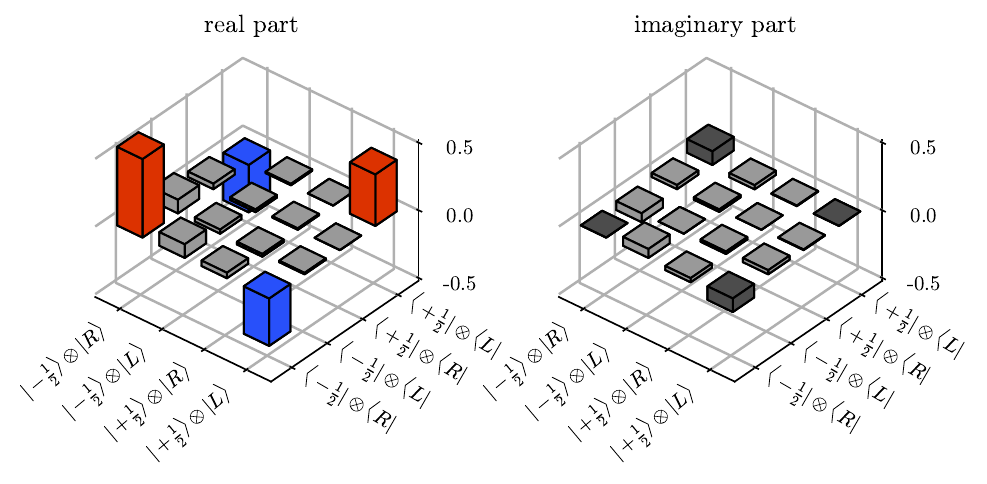}
	\caption{Reconstructed ion-photon density matrix with a fidelity of $83(2)$\,\% including background correction.}
 \label{fig:IonPhotonExperimentsEnt}
\end{figure}

\begin{figure}[htb]
	\hfill
	\begin{minipage}[t]{4.25cm}
		\includegraphics[width=4.25cm]{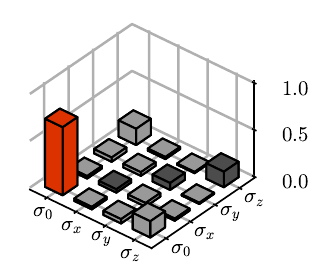}
		
		(a)\hspace{2mm} $\ket{\Phi^-}$
	\end{minipage}
	\hfill
	\begin{minipage}[t]{4.25cm}
		\includegraphics[width=4.25cm]{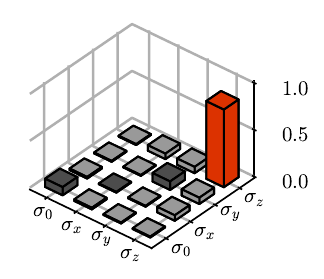}
		
		(b)\hspace{2mm} $\ket{\Phi^+}$\vspace{0.5cm}
		\mbox{}
	\end{minipage}
	\hfill\mbox{}.
	\hfill
	
	\begin{minipage}[t]{0.5cm}
		(c)
	\end{minipage}
	\hfill
	\begin{minipage}[t]{8cm}
		\begin{tabular}{c | c | c}
			state & fidelity w correction & fidelity w/o correction \\
			\hline
			$\ket{\Phi^-}$ & 80(6)~\% & 78(6)~\%\\
			$\ket{\Phi^+}$ & 87(5)~\% & 86(5)~\%\\
		\end{tabular}
	\end{minipage}
	\caption{Ion-to-photon quantum state teleportation over the fiber link. (a) and (b): reconstructed process matrices of the teleportation conditioned on the Bell-state measurement results $\ket{\Phi^-}$ and $\ket{\Phi^+}$ (background and binning correction is applied). (c) Summarized fidelities of the ion-to-photon teleportation for the two Bell measurement results, with and without correction.}
	\label{fig:IonPhotonExperimentsTele}
\end{figure}

The pair source is pumped with $15$\,mW for both protocols, leading to an initial background-corrected fidelity of $\bra{\Psi^+}\rho\ket{\Psi^+} = 98.0(1)$\,\% between one $854$\,nm photon detected at the UdS and its $1550$\,nm partner photon detected at the HTW. Without background correction, a fidelity of $83.6(3)$\,\% is measured, averaged over $\approx14$ days of measurement time. 
The deviation to the measurement of figure \ref{fig:PPEntWStabilization} is mainly attributed to a phase drift of the photon pair state and a minor contribution of accumulated polarization drifts in the unstabilized fiber parts where the polarization calibration is applied every 30\,min respective 2\,h (see previous section). The protocols use an exposure time of the ion of $400\,\si{\micro\second}$ for photon absorption, and the spin-echo technique described in \cite{Arenskoetter2024} is implemented to extend the atomic coherence time. 

The result for the ion-photon entanglement protocol is shown in figure~\ref{fig:IonPhotonExperimentsEnt}. It displays the tomographically reconstructed 2-qubit density matrix of the photon at the HTW and the memory qubit, after the state mapping. The basis for the memory qubit are the Zeeman states $\ket{\pm\frac{1}{2}} = \ket{\text{S}_{1/2},m=\pm\frac{1}{2}}$, for the photonic qubit the circular polarization basis $\ket{R/L}$ is used. The fidelity with the ideal state 
\begin{equation}
    \ket{\Phi}_\mathrm{at,ph} = \frac{1}{\sqrt{2}} \left( \ket{-\frac{1}{2}}\otimes \ket{R} - \ket{+\frac{1}{2}} \otimes \ket{L} \right)
\end{equation}
is $79(2)$\,\% ($83(2)$\,\% with background correction), the state purity is $71(3)$\,\% ($79(4)$\,\% with background correction). Comparing this result to the direct generation of ion-photon entanglement obtained by our group \cite{Bock2018} with $98.2(2)\,\%$ fidelity, one clearly observes a reduction. This reduction must be attributed to the aggregated complexity due to the combination of the SPDC source as additional component and external resource of entanglement with the atomic setup which is operated in heralded-receiver mode. The increased complexity requires longer integration times and hence higher requirements on longterm stability of the individual setups. Additionally a tradeoff between success probability and atomic decoherence must be made by the selection of the exposure window. In conclusion, if only atom-photon entanglement is required, then the protocol of \cite{Bock2018} is advantageous. On the other hand, heralded absorption allows us to implement quantum state teleportation, as described in the following.

Figure \ref{fig:IonPhotonExperimentsTele}(a) and (b) shows the reconstructed process matrices of the quantum teleportation protocol. The process matrix describes how the protocol maps the prepared memory qubit onto the photonic qubit at the HTW in terms of the Pauli operators $\{\sigma_0, \sigma_x, \sigma_y, \sigma_z\}$ \cite{Chuang1997}. It thereby indicates which local operation to the photon would be required to restore the initial memory state. Two of the four Bell states are distinguished by the BSM, and quantum process tomography is performed separately for each measurement result. One can see that no rotation is necessary when the projection results in the $\ket{\Phi^-}$ Bell state (figure\,\ref{fig:IonPhotonExperimentsTele}(a)), while a $\sigma_z$ rotation is required when the result is $\ket{\Phi^+}$ (figure\,\ref{fig:IonPhotonExperimentsTele}(b)). This corroborates the success of the teleportation protocol. We identify 78(6)\,\% and 86(5)\,\% as the process fidelity, corresponding to the diagonal element of the process matrix for the expected rotation, $\sigma_0$ or $\sigma_z$. The fidelity values including background correction are summarized in table \ref{fig:IonPhotonExperimentsTele}(c). This process fidelity compares well to the measured $86(7)\,\%$ published in \cite{Arenskoetter2024}, and shows that the additional complexity due to frequency conversion and the use of the deployed fiber is not affecting the application.

\section{Conclusion}\label{sec:Conc}
To conclude, we investigated a 14.7\,km deployed fiber link that connects two nodes across the urban area of Saarbr\"ucken, with respect to its application for quantum network protocols. We characterized general properties such as loss, photonic background noise, phase stability and polarization-dependent loss. We showed that this fiber faithfully transmits polarization-encoded qubits at a process fidelity level of 99\,\% for $\sim 100$\,s without correction. To use the link repeatedly for longer times at the same level of fidelity, we implemented polarization stabilization in a fully automated way with a reasonable duty cycle between transmission and stabilization of $\approx100\,\text{s}/1.1\,\text{s} \approx 91$. We then used a type-\rom{2} SPDC entangled photon pair source, a $^{40}$Ca$^+$ single-ion quantum memory, and quantum frequency conversion to realize elementary quantum communication protocols over the link, such as entanglement distribution, atom-photon entanglement, and quantum state teleportation. 

Entanglement distribution over the link exhibits fidelities $\mathcal{F}>98\,\%$ for up to $60\,$s, underlining the suitability of the implemented methods for establishing a quantum communication channel. Ion-photon entanglement between the memory qubit and a converted and transmitted photon in the telecom C-band was demonstrated with $\mathcal{F}=79(2)\,\%$ fidelity, employing heralded absorption of one photon of a SPDC pair. By utilizing the heralded absorption as a Bell state measurement, we also realized quantum state teleportation with fidelities of 86(5)\,\% and 78(6)\,\%. These proof-of-concept experiments highlight the potential of the single-ion platform, in combination with QFC, for implementation of quantum networks over deployed fiber testbeds; they pave the way to future experiments, such as the implementation of QKD protocols and remote memory entanglement for quantum repeaters. The fiber link also lends itself to the application of calibration-independent certification, following the example of \cite{Bock2023}. 

Open questions, to be addressed in future research on the fiber link, are the origin and time dependence of the polarization-dependent loss, as well as the origin, and possible compensation, of phase fluctuations in the fiber transmission, which are particularly relevant for interference-based quantum communication protocols.

\section*{Author contributions}
S.K., C.H., E.A., and J.M. designed, performed, and analyzed the trapped ion, pair source, polarization characterization, and polarization stabilization parts of the experiments, T.B. the QFC part. M.S., T.B. and R.B., M.Y, M.L., R.H. from Menlo Systems performed the interferometric length change characterization. S.K. and C.H. wrote the paper with input from all authors. J.E. and C.B. planned and supervised the project.

\begin{acknowledgments}
We acknowledge support by the German Federal Ministry of Education and Research (BMBF) through projects Q.Link.X (16KIS0864), QR.X (16KISQ001K) and HiFi (13N15926). 

We are grateful to Damian Weber from the HTW saar for providing necessary lab infrastructure, to Dieter Klein and colleagues of HIZ Saarland for indispensable support with the fiber link and communication infrastructure, and to VSE NET for generously providing the fiber link itself. 
\end{acknowledgments}

\appendix
\section*{Appendix}
\section{Comments on polarization-dependent loss}\label{appendix: PDL}

We treat the polarization qubit, according to \cite{Nielsen2000}, as a \textit{dual-rail} qubit in two orthogonal polarization modes,
\begin{equation}
   \ket{\varphi} = \left(\alpha \ a^\dagger_\text{P}  + \beta \ a^\dagger_{\text{P}_\perp}\right) \ket{vac}\ \ ,
\end{equation}
with $|\alpha|^2 + |\beta|^2 = 1$. Polarization-dependent loss (PDL) is then treated as a partial beam splitter effecting the unitary transformation 
$\mathds{1}_{\text{P}} \otimes \text{exp}[\vartheta (a^\dagger_{\text{P}_\perp}b_{\text{P}_\perp} + a_{\text{P}_\perp}b^\dagger_{\text{P}_\perp})]$, coupling  one polarization mode to the environment. Here $a^\dagger_{\text{P}}$, $a^\dagger_{\text{P}_\perp}$, $a_{\text{P}_\perp}$ are the creation and annihilation operators of the transmitted modes, and 
$b^\dagger_{\text{P}_\perp}$, $b_{\text{P}_\perp}$ are the creation and annihilation operators of the environment mode. The transmission amplitude is given by $T=\cos(\vartheta)$. In this sense, every loss process in the transmission channel couples the state to a new spatial mode from the environment. The post-selective nature of our experiments projects onto the transmitted fraction. It follows that polarization-dependent loss is described by a non-trace-preserving process. 
\begin{figure*}[t]
\includegraphics[width=\textwidth]{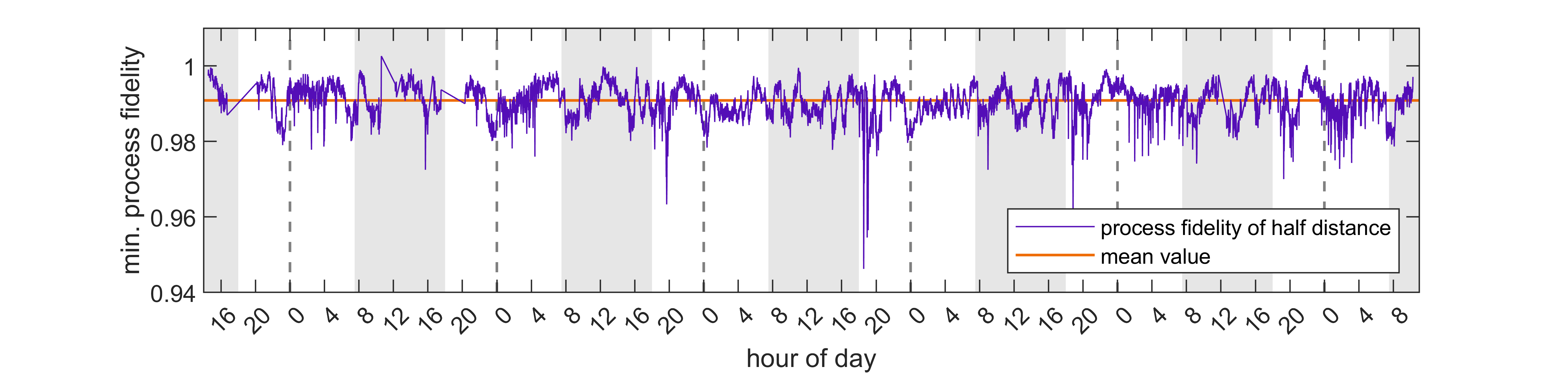}
\caption{Time trace of the minimum process fidelity (purple) and the mean value (red). }
\label{fig:PDL-PF}
\end{figure*}
We now change the representation from the general \textit{dual-rail} qubit to the polarization basis $\{\ket{0},\ket{1}\}$ and write the state that is unaffected by the polarization-dependent loss as
\begin{equation}
    \ket{\text{P}} = e^{-i\,\Phi/2} \cos{\nicefrac{\Theta}{2}} \ket{0} + e^{i\,\Phi/2} \sin{\nicefrac{\Theta}{2}} \ket{1} ~~,
\end{equation}
using the angles $\Phi$ and $\Theta$ of the Bloch sphere representation. 
The loss operator describing PDL is then 
\begin{equation}
\hat{B} = \ket{\text{P}}\bra{\text{P}} + \text{T}\ket{\text{P}_\perp}\bra{\text{P}_\perp} = \text{T}\,\mathds{1} + (1-\text{T}) \ket{\text{P}}\bra{\text{P}}
\label{appendix:eq:PDL-M}
\end{equation} 
with the orthogonal state $\ket{\text{P}_\perp}$ ($\braket{\text{P}|\text{P}_\perp}=0$) and the transmission amplitude T. The action of the PDL on an arbitrary input state, described by the density matrix $\hat{\rho}_{in}$, is then represented by a quantum process 
$\varepsilon\left(\hat{\rho}_{in}\right)$; including normalization, this results in an output density matrix
\begin{align}
    &\hat{\rho}_{out} = \frac{\varepsilon\left(\hat{\rho}_{in}\right)}{\text{tr}\left(\varepsilon\left(\hat{\rho}_{in}\right)\right)}= \frac{\hat{B}\,\hat{\rho}_{in}\hat{B}^\dagger}{\text{tr}\left(\hat{B}\,\hat{\rho}_{in}\hat{B}^\dagger\right)}\label{appendix:eq:PDL-Mapplied}\\
    &\ =\frac{\left(\text{T}\,\mathds{1} + (1-\text{T}) \ket{\text{P}}\bra{\text{P}}\right)\ \hat{\rho}_{in}\ \left(\text{T}\,\mathds{1} + (1-\text{T}) \ket{\text{P}}\bra{\text{P}}\right)}
    {\text{T}^2\ \bra{P_\perp}\hat{\rho}_{in}\ket{P_\perp} + \bra{P}\hat{\rho}_{in}\ket{P}}\nonumber\\
    &\ =\frac{\left( \frac{1+\text{T}}{2}\,\mathds{1} +  \frac{(1-\text{T})\,\Vec{\Gamma}}{2\,\gamma} \Vec{\hat{\sigma}}\right)\ \hat{\rho}_{in}\ \left(\frac{1+\text{T}}{2}\,\mathds{1} +  \frac{(1-\text{T})\,\Vec{\Gamma}}{2\,\gamma} \Vec{\hat{\sigma}}\right)}
    {\text{T}^2\ \bra{P_\perp}\hat{\rho}_{in}\ket{P_\perp} + \bra{P}\hat{\rho}_{in}\ket{P}}\label{appendix:eq:PDL-Mpauli}\\
    &\ = \frac{\mathds{1} + \overbrace{\frac{ \sqrt{1-\gamma^2} \Vec{\lambda}_{in} 
    + \left( \frac{1-\sqrt{1-\gamma^2}}{\gamma^2}\ \Vec{\lambda}_{in} \, \Vec{\Gamma} + 1 \right)\ \Vec{\Gamma}}{1+\Vec{\lambda}_{in} \, \Vec{\Gamma}}}^{\Vec{\lambda}_{out} } \ \Vec{\hat{\sigma}}}{2}\label{appendix:eq:PDL-lambdaout}
\end{align}
 using the Pauli vector $\Vec{\hat{\sigma}} = [\hat{\sigma}_1,\hat{\sigma}_2,\hat{\sigma}_3]$ and the transformation to the Bloch vectors by 
\begin{align}
    \hat{\rho}_{in} &= \frac{\mathds{1} + \Vec{\lambda}_{in} \Vec{\hat{\sigma}}}{2}\\
    \ket{P}\bra{P} &= \frac{\mathds{1} + \frac{\Vec{\Gamma}}{\gamma} \Vec{\hat{\sigma}}}{2}\ \ \ \text{with} \ \ \ \gamma = |\Vec{\Gamma}| = \frac{1-\text{T}^2}{1+\text{T}^2}\label{appendix:eq:PDL-gamma}\\
    \bra{P}\hat{\rho}_{in}\ket{P} &= \frac{1 + \Vec{\lambda}_{in}\,\frac{\Vec{\Gamma}}{\gamma}}{2}\\
    \bra{P_\perp}\hat{\rho}_{in}\ket{P_\perp} &= \frac{1 - \Vec{\lambda}_{in}\,\frac{\Vec{\Gamma}}{\gamma}}{2}
\end{align}
The PDL vector $\Vec{\Gamma}$ is parallel to the Bloch vector of the unaffected state ($\ket{\text{P}} \bra{\text{P}}$) with the norm containing the transmission according to equation \ref{appendix:eq:PDL-gamma}. Using this notation, the output Bloch vector $\Vec{\lambda}_{out}$ of equation \ref{appendix:eq:PDL-lambdaout} is equal to the description of \cite{Gisin1995}, now derived starting from the quantum mechanical description of a beam splitter.  \cite{Gisin1995} shows that concatenation of two PDL vectors can again be experessed by PDL vector, which allows to describe the whole transmission link as a single PDL vector, independent how the loss is compound along the link. But it is not obvious to us that the description of PDL as a simple beam-splitter operation is correct in general. If, for example, the origin of the polarization-state dependent loss is scattering, then also a state dependent phase shift must be included in the description. This aggravates the description of a concatenated transmission link. It is not obvious if this can be decomposed in a rotary- and a PDL-part or if it must be described position dependent.
To this end, by decomposing the nominator of equation \ref{appendix:eq:PDL-Mpauli} into the Pauli basis $\{\sigma_0,\sigma_1,\sigma_2,\sigma_3\}$ allows to set a lower bound to the process fidelity. The process matrix $\chi$ defined by
\begin{equation}
    \varepsilon\left(\hat{\rho}_{in}\right) = \sum_{i,j=0}^4 \chi_{i,j} \sigma_i \hat{\rho}_{in} \sigma_j^\dagger \ \ .
\end{equation}
can directly be identified to
\begin{equation}
    \chi = 
\begin{pmatrix}
\nicefrac{(1+\text{T})}{2}\\
\nicefrac{(1-\text{T})\,\Vec{\Gamma}}{(2\gamma)}
\end{pmatrix}
\begin{pmatrix}
\nicefrac{(1+\text{T})}{2} & \nicefrac{(1-\text{T})\,(\Vec{\Gamma})^\top}{(2\gamma)}
\end{pmatrix}
\end{equation}
from the nominator, and the maximum of 1 of the denominator, i.e. the maximum norm, bounds the process fidelity to 
\begin{equation}
    \mathcal{F}_{\text{P}} \geq \frac{(1+\text{T})^2}{4}
\end{equation}
Figure \ref{fig:PDL-PF} shows the time trace and the mean minimum process fidelity calculated from the measured PDL of figure \ref{fig:PDL}(a).
The PDL of each fiber is inferred as half of the total measured PDL as described in the main text. One can see that, despite a few spikes, the minimum stays well above 98\,\%. The spikes are attributed to occasionally occurring stress that is applied onto the fiber. We can only guess for their origins such as maintenance on the patch stations, gusting wind at the overhead cable, heavy vehicles stressing the underground cable or a more careful description of PDL must be investigated.

\section{Polarization stabilization algorithm}
\label{appendix:PolStabAlgorithm}
In the following, the polarization stabilization scheme is explained in more detail. A Gradient-Descent algorithm is used to minimize a real, differentiable error function, $f(\vec{U})$, that measures the deviation from the target polarization state. The parameter space is given by four voltages applied to the piezo polarization controller, $\vec{U}=(U_1,U_2,U_3,U_4)$. The scheme only corrects for polarization rotations, therefore two input polarizations are sufficient to fully define the process. Here, horizontal and diagonal linear polarizations with Stokes-vectors $\vec{S}_{H}$ resp. $\vec{S}_{D}$ are injected at the sender, and polarizations $\vec{S}_1(\vec{U})$ resp. $\vec{S}_2(\vec{U})$ are measured at the receiver, as functions of the applied piezo voltages $\vec{U}$. The error function is then computed by
\begin{equation}\label{eq:StabErrorFun}
	f(\vec{U}) =\|\vec{S}_1(\vec{U})-\vec{S}_{H}\|_2^2+\|\vec{S}_2(\vec{U})-\vec{S}_{D}\|_2^2~.
\end{equation}  
The $i$-th component of the gradient is now calculated by varying the voltage of each piezo element by the search voltage step size $\Delta U$ in two directions:
\begin{align}\label{eq:StabErrorFunGradient}
	(\Delta f(\vec{U}))_i &=\frac{f(\vec{U}-\vec{e}_i\,\Delta U)-f(\vec{U}+\vec{e}_i\,\Delta U)}{2 |\Delta U|},
\end{align}  
where $\vec{e}_i$ are the unit vectors of the four-dimensional parameter space. The new set of piezo voltages $\vec{U}_\text{new}$ is then obtained by updating the last set $\vec{U}_\text{old}$ to
\begin{align}
	\vec{U}_\text{new} = \vec{U}_\text{old} - D \Delta f(\vec{U}_\text{old}),
\end{align}  
using stepsize $D$. After each step, the process fidelity $\mathcal{F}_\text{P}$ according to eqs.~(\ref{eq:ProcessFidelity}) and (\ref{eq:traceM}) is measured. The algorithm terminates once $\mathcal{F}_\text{P}$ reaches a predetermined threshold value, $\mathcal{F}_\text{P,th}$. The stepsize $D$, as well as the search voltage $\Delta U$ are kept constant at initial values $D_0$ and $\Delta U_0$ until a process fidelity of $\mathcal{F}_\text{P,0}=95$\,\% is reached. Upon reaching this, the two parameters are adjusted according to 
\begin{align}
	D&=\frac{1-\mathcal{F}_\text{P}}{1-\mathcal{F}_\text{P,0}}D_0+D_1 \ \ ,\\
	\Delta U&=\frac{1-\mathcal{F}_\text{P}}{1-\mathcal{F}_\text{P,0}}\Delta U_0+\Delta U_1
\end{align}
with the minimum stepsizes $D_1$ and $\Delta U_1$. This ensures that the discretization does not affect the search for the optimal point of process fidelities. The parameters $D_0, D_1, \Delta U_0, \Delta U_1$ are determined empirically. Minimizing the error function $f(\vec{U})$ is equivalent to maximizing the process fidelity $\mathcal{F}_\text{P}$ under the assumption that polarization rotations are the only error source. The error function could also be defined with the process fidelity directly, but the numerical effort to calculate the gradient, eq.~(\ref{eq:StabErrorFunGradient}), would become larger. 

\bibliography{Bibliography.bib}
\end{document}